\documentclass[prd, nofootinbib, aps, 10pt, notitlepage]{revtex4-1}
\pdfoutput=1

\usepackage{newtxmath}
\usepackage{newtxtext}
\usepackage{bm}
\usepackage{graphicx} 
\usepackage{epsfig}
\usepackage{subfigure}
\usepackage{dcolumn} 

\usepackage[usenames,dvipsnames]{color}
\usepackage[pagebackref=false, colorlinks=true]{hyperref}
\definecolor{redish}{rgb}{0.7,0.2,0.0}  
\definecolor{bluish}{rgb}{0.2,0.5,0.8}
\usepackage{soul}

\hypersetup{linkcolor=redish,     
                  citecolor=blue,        
                  filecolor=magenta,   
                  urlcolor=bluish}        

\DeclareFontFamily{U}{rsfs}{}         
\DeclareFontShape{U}{rsfs}{m}{n}{<5> rsfs5 <6><7> rsfs7          %
  <8><9><10><10.95><12><14.4><17.28><20.74><24.88> rsfs10}{}     %
\DeclareMathAlphabet{\mathfs}{U}{rsfs}{m}{n}

\begin{document}

\title{An Approach to Stability Analyses in General Relativity via Symplectic Geometry}
\author{Prashant Kocherlakota}
\email{k.prashant@tifr.res.in}
\affiliation{Tata Institute of Fundamental Research, Mumbai 400005, India}
\author{Pankaj S. Joshi}
\email{psjprovost@charusat.ac.in}
\affiliation{International Center for Cosmology, CHARUSAT University, Anand 388421, India}

\begin{abstract}
We begin with a review of the statements of non-linear, linear and mode stability of autonomous dynamical systems in classical mechanics, using symplectic geometry. We then discuss what the Arnowitt-Deser-Misner (ADM) phase space and the ADM Hamiltonian of general relativity are, what constitutes a dynamical system, and subsequently present a nascent attempt to draw a formal analogy between the notions of stability in these two theories. We wish to note that we have not discussed here the construction of the reduced phase space by forming the quotient space of the constrained phase space with the gauge orbits. Our approach here is pedagogical and geometric, and the motivation is to unify and simplify a formal understanding of the statements regarding the stability of stationary solutions of general relativity. That is, typically the governing equations of motion of a Hamiltonian dynamical system are simply the flow equations of the associated symplectic Hamiltonian vector field, defined on phase space, and the non-linear stability analysis of its critical points have simply to do with the divergence of its flow there. Further, the linear stability of a critical point is related to the properties of the tangent flow of the Hamiltonian vector field.  

In the second half of this work, we posit that a study of the genericity of a particular black hole or naked singularity spacetime forming as an endstate of gravitational collapse is equivalent to an inquiry of how sensitive the orbits of the symplectic Hamiltonian vector field of general relativity are to changes in initial data. We demonstrate this by conducting a restricted non-linear stability analysis of the formation of a Schwarzschild black hole, working in the usual initial value formulation of general relativity. 
\end{abstract}

\maketitle


\section{Introduction}
In the framework of general relativity (GR), spacetime is modelled by a 3+1-dimensional Lorentzian manifold equipped with a metric, which is a smooth, non-degenerate, symmetric two-tensor field. The governing dynamical equations of GR are the Einstein field equations (EFEs), which relate the local spacetime curvature, characterized by the metric tensor, with the local energy and momentum within that spacetime, described by a stress--energy tensor. Various well known stationary solutions of the EFEs describe geometries of spacetimes that contain curvature singularities \cite{Schwarzschild16, Reissner16, Nordstrom18, Taub51, Newman+63, Kerr63, Newman+65, Janis+68, Hawking_Ellis73}, and are of interest here. Depending on whether or not they are completely shielded from an asymptotic observer by an event horizon, these spacetimes are classified as containing either a black hole or a (globally visible) naked singularity respectively \cite{Joshi93}. Understanding how such novel objects form from gravitational collapse (existence), the genericity of their occurrence, their stability when they form and what observational features these objects would exhibit have been long standing sources of intrigue in the theoretical GR community. While black hole spacetimes possess a plethora of interesting physical features like event horizons, ergoregions, photon spheres etc., the existence of a globally visible naked singularity represents the unique possibility of an observable extremely high-curvature, strong-gravity region. Naturally, such compact objects are sources of great astrophysical interest as well, with various large-scale astronomy missions like the (\href{https://www.lsc-group.phys.uwm.edu/ppcomm/Papers.html}{LIGO}; \cite{Abbott+16}) and the Virgo interferometer \cite{Accadia+12}, and the next generation Laser Interferometric Space Antenna (\href{https://www.lisamission.org/news/Papers}{LISA}; see for example \cite{Amaro-Seoane+07}), the Event Horizon Telescope (\href{https://eventhorizontelescope.org/}{EHT}; \cite{Doeleman+08, Doeleman+12, Akiyama+19}), the Square Kilometre Array (\href{https://www.skatelescope.org/}{SKA}; \cite{Braun+14}) and \textit{Fermi} \cite{Atwood+09} all actively looking to detect and study them.

Various instances of black holes and naked singularities forming as a result of gravitational collapse have been reported \cite{Datt38, Oppenheimer_Snyder39, Lemaitre33, Tolman34, Bondi47, Vaidya66, Joshi+11}, and questions regarding the genericity of their occurrence have been explored extensively. Results like the Birkhoff theorem \cite{Birkhoff23} and various other analytical \cite{Sasaki_Nakamura90, Dafermos_Rodnianski10, Lucietti_Reall12, Duztas_Semiz13, Dafermos+14, Duztas15, Shlapentokh-Rothman15, Natario+16, Richartz16} and numerical studies \cite{Sasaki_Nakamura82, Miller_Motta89, Yo+02, Baiotti+05, Nathanail+17} have bolstered our expectation that black holes do, in fact, form as generic endstates of continual gravitational collapse. However, despite various important efforts \cite{Christodoulou84, Christodoulou86, Ori_Piran87, Ori_Piran90, Shapiro_Teukolsky92, Joshi_Dwivedi93, Choptuik93, Christodoulou94, Christodoulou99b, Harada+02, Crisford_Santos17}, the genericity of evolutions of regular configurations of matter to naked singularities is less clear. That is, even if naked singularities do form for some particular choice of initial data of gravitational collapse $d$ in the space of all allowed initial data $\mathcal{D}$, do they also form for other arbitrary initial data chosen within an infinitesimal neighborhood of $d$ in $\mathcal{D}$? Further, how is this space of all allowed initial data $\mathcal{D}$ partitioned into regions that produce black holes versus those that generate naked singularities? These are fundamental open questions in GR and are related to the cosmic censorship hypotheses \cite{Penrose65, Penrose69, Penrose79, Israel84, Israel86, Christodoulou99a}, which roughly require a positive or negative statement regarding whether the set of initial data that eventually lead to naked singularities make up a `sufficiently significant' portion of the space of all allowed initial data. In \S\ref{sec:Stab_GC}, we will argue that when one inquires about the genericity of the formation of black holes or naked singularities (or any other astrophysical objects), one is in fact concerned about the non-linear stability of their formation processes against changes in initial data. This statement will become clear in \S\ref{sec:OSD_Non-Linear} when we consider the evolution of different initial configurations of pressureless fluids to either black holes or naked singularities via the the Lema{\^i}tre-Tolman-Bondi \cite{Lemaitre33, Tolman34, Bondi47} collapse solutions of GR.

The study of the stability of these objects when they do form is naturally of fundamental importance in GR and theoretical astrophysics. For example, unless small `disturbances' constrained to the exterior of a black hole or a naked singularity oscillate around the background metric without causing significant deviations in it, in the same way a mechanical system oscillates around a local minimum configuration of its potential energy, these solutions of GR would be astrophysically uninteresting (since they would imply that these objects are short-lived and therefore difficult to observe). 
One can also ask, in similar vein, whether more exotic objects like bose stars \cite{Kolb_Tkachev93}, preon stars \cite{Hansson_Sandin05} or superspinars \cite{Gimon_Horava09} can be stable. 

Recognizing the importance of stablity analyses of spacetimes in general relativity, a study of mode stability from the point of view of metric perturbations was attempted in a seminal paper by Regge \& Wheeler \cite{Regge_Wheeler57}. Since the Kruskal-Szekeres extension of the Schwarzschild spacetime \cite{Kruskal60, Szekeres60} was not yet discovered at the time, the coordinate singularity at the Schwarzschild horizon made it difficult to ascertain whether or not divergences exhibited by perturbation at this surface were real. This study was eventually completed by Vishveshwara \cite{Vishveshwara70}, who showed that the Schwarzschild spacetime is indeed mode stable. Zerilli \cite{Zerilli70} unified the discussion of stability analyses for Regge-Wheeler's even and odd parity perturbations by discovering a transformation that connected the potentials for these two classes and a gauge-invariant formulation of metric perturbations was later provided by Moncrief \cite{Moncrief75}. An alternative approach via the Newman-Penrose \cite{Newman_Penrose62} formalism was set up by Bardeen \& Press \cite{Bardeen_Press73} for the mode analysis of the Schwarzschild spacetime, which was later extended to the Kerr family by Teukolsky \cite{Teukolsky73}, and the celebrated result regarding the mode stability of Kerr black holes was obtained in the seminal paper of Whiting \cite{Whiting89}. Various authors have since studied the stability of Kerr naked singularities \cite{Dotti+08}, Kerr superspinars \cite{Pani+10, Nakao+18} and other objects \cite{Cunningham+78, Cunningham+79, Cunningham+80, Kokkotas_Schmidt99, Bini+03, Berti+04, Cardoso+08, Berti+09, Aretakis11a, Aretakis11b, Aretakis13, Toth16}.

Statements regarding the various notions of stability in GR are typically made in the initial value formulation of GR and there are several excellent review articles in the literature on the same \cite{Rendall05, Rodnianski06, Christodoulou09, Ringstrom09, Dafermos_Rodnianski13, Dafermos14}. We propose here that restating these notions using symplectic geometry earns us substantial insight since one can then draw formal analogies between the seemingly abstruse notions of stability of spacetimes in general relativity and the more familiar notions of stability in classical Galelei-Newton mechanics. 

The organisation of this paper is as follows. In \S\ref{sec:Stab_CM}, we begin with a review of the phase space of classical mechanics (CM), how Hamiltonian flows defined on it and the notions of stability of classical mechanical systems. In \S\ref{sec:Stability_GR}, we begin by discussing the initial value formulation of general relativity and how valid initial data sets are defined. Then we briefly touch upon what the phase space and Hamiltonian of GR are, and attempt to draw a formal analogy to discuss the notions of stability of spacetimes in GR.  As mentioned above, in \S\ref{sec:Stab_GC} we demonstrate how, reverting back to the usual initial value formulation of GR, a typical non-linear stability analysis proceeds in GR. There we consider the class of Lema{\^i}tre-Tolman-Bondi \cite{Lemaitre33, Tolman34, Bondi47} dust collapse models, of which the Oppenheimer-Snyder-Datt (OSD, \cite{Datt38, Oppenheimer_Snyder39}) collapse model is a member. The significance of this particular study is that the OSD collapse process terminates in the formation of a Schwarzschild black hole, and we are essentially analysing how changes in the initial data from which the collapse evolves (under Einstein's field equations) affect its formation. 

\section{A Geometric Approach to Stability in Classical Mechanics} \label{sec:Stab_CM}
In this section, we review the formal statements of non-linear, linear and mode stability of dynamical systems in classical Galelei-Newton mechanics. We begin with a quick summary of Hamilton's equations and in \S\ref{sec:Classical_Dynamics_Symplectic_Geometry}, we restate their dynamical content in the language of symplectic geometry. In specific, Hamilton's equations can equivalently be thought of as being the flow equations of an appropriately defined symplectic Hamiltonian vector field, with trajectories in phase space corresponding to flows of this vector field. We present the construction of the phase space of a classical mechanical system, which is a symplectic manifold, discuss what constitutes a Hamiltonian system and introduce the concept of the flow of the aforementioned Hamiltonian vector field. Since our eventual aim is to discuss the various formulations of stability analyses of such dynamical systems, we will also define `tangent (to the Hamiltonian vector field) flows.' In \S\ref{sec:Stability_CM}, we discuss the various notions of stability and it will become apparent that this geometric approach supplies valuable insight. In \S\ref{sec:Stability_GR}, we will transport the geometric intuition acquired here into the context of general relativity via a formal analogy.

Our discussions here will be limited to autonomous or time-independent Hamiltonian dynamical systems, whose descriptions on symplectic manifolds is well established \cite{Abraham_Marsden78, Arnold80, Bryant95, Jose_Saletan98, Marsden_Ratiu99, Frankel01, Farantos14, Apostol74}. We take the view that a discussion of non-autonomous Hamiltonian systems does not add substantial additional insight towards our primary goal of highlighting the analogy between the notions of stability in CM and GR. Further, noting that non-autonomous Hamiltonian dynamical systems can, in various settings, be replaced by autonomous Hamiltonian dynamical systems defined on `extended phase space' (see for example \cite{Struckmeier_Riedel00, Struckmeier05}), we shall conveniently omit a review of such systems.

Now, in the canonical Hamiltonian formulation of CM, an arbitrary instantaneous state of a dynamical system with $n$-degrees of freedom is characterized by specifying its generalised coordinates $(q^1, q^2, \cdots, q^n)$ and momenta $(p_1, p_2, \cdots, p_n)$. The collection of all such possible states is called its phase space. Then, given a Hamiltonian function $H(q^1, q^2, \cdots, q^n, p_1, p_2, \cdots, p_n)$ defined over phase space, the governing dynamical equations are the Hamilton equations given as,
\begin{equation} \label{eq:Hamilton_Equations}
\frac{dq^i}{dt} = \frac{\partial H}{\partial p_i},\ \frac{dp_i}{dt} = -\frac{\partial H}{\partial q^i}.
\end{equation}
That is, given a specific initial state $(q^1(0), q^2(0), \cdots, q^n(0), p_1(0), p_{2}(0), \cdots, p_{n}(0))$ of the system, its unique future time evoution is obtained by solving the initial value problem of Hamilton's equations.

\subsection{Dynamics using Symplectic Geometry} \label{sec:Classical_Dynamics_Symplectic_Geometry}
The instantaneous configuration of an autonomous dynamical system, in classical mechanics, is described by the values of the $n$-generalized coordinates $(q^1, q^2, \cdots, q^n)$ and corresponds to a particular point $q$ in configuration space $Q$. An element in the cotangent bundle $T^*Q$ of configuration space consists of a 1-form defined in the cotangent space $T_q^*Q$ at every point $q\in Q$, and such a form is given by its $n$-components $(p_1, p_2, \cdots, p_n)$, which characterize the instantaneous generalized momentum of the system of interest (see for example Ch. 8 of \cite{Arnold80}). These $2n$ numbers, denoted succinctly by $z^a \equiv (q^1, q^2, \cdots, q^n, p_1, p_2, \cdots, p_n)$, with the index  $a$ running from $1$ to $2n$, form a collection of local coordinates
for points in $T^*Q$, which can then immediately be identified as the momentum phase space (henceforth, just phase space) of the dynamical system under consideration. 

Further, since phase space is the cotangent bundle of a smooth manifold, it comes naturally equipped with a symplectic structure and is therefore a symplectic manifold; i.e., there exists a closed, non-degenerate differential 2-form $\omega$ on $T^*Q$. In the current context, this symplectic 2-form is simply given as, 
\begin{equation}
\omega = \sum_{i=1}^{n}dp_i \wedge dq^i.
\end{equation}
Since $\omega$ is non-degenerate, it sets up an isomorphism from the tangent bundle to the cotangent bundle of phase space, $\omega: T\Pi \rightarrow T^*\Pi$, where we have introduced $\Pi \equiv T^*Q$ to denote phase space. Then for every smooth, real-valued function $H: \Pi \rightarrow \mathbb{R}$, one can associate a unique vector field $X_H$ via $\omega(X_H, \bullet) = -\text{d}H$. Introducing the inverse isomorphism, $\Omega: T^*\Pi \rightarrow T\Pi,\ \Omega = \omega^{-1}$, which in local coordinates is given as,
\begin{equation}
\Omega^{ab} = 
\begin{bmatrix}
0_n & I_n \\
-I_n & 0_n
\end{bmatrix} ,
\end{equation}
where $O_n$ and $I_n$ are the $n$-dimensional zero and identity matrices, we find $X_H \equiv \Omega~\text{d}H$. And in local coordinates we can write,
\begin{equation}
 X^a_H = \Omega^{ab}\partial_b H = \Omega^{ab}\frac{\partial H}{\partial z^b} = \left(\frac{\partial H}{\partial p_1}, \frac{\partial H}{\partial p_2}, \cdots, \frac{\partial H}{\partial p_n}, - \frac{\partial H}{dq^1}, -\frac{\partial H}{dq^2}, \cdots, -\frac{\partial H}{dq^n}\right).
\end{equation}
$X_H$ is a symplectic vector field since it leaves the symplectic structure invariant, i.e. it satisfies $\mathcal{L}_{X_H}\omega = 0$, where $\mathcal{L}_{X_H}$ is the Lie-derivative w.r.t. $X_H$. Any such function $H$ is called a Hamiltonian function and $X_H$ is the associated symplectic Hamiltonian vector field.

The flow of $X_H$ is a 1-parameter Lie group of symplectomorphisms $\phi_H^t: \mathbb{R} \times \Pi \rightarrow \Pi$ such that for some $z_0 \in \Pi$, $\phi^t_H(z_0): \mathbb{R} \rightarrow \Pi$ is an orbit or integral curve of $X_H$, which passes through $z_0$ at $t=0$. Let us denote such an orbit in local coordinates as $z^a(t) \equiv z^a(\phi^t_H(z_0))$. Then we can write,
\begin{equation} \label{eq:Hamiltonian_Flow}
\dot{z}^a(t) = X^a_H\left(z(t)\right),\ z^a(0) = z_0^a,
\end{equation}
where the overdot represents differentiation w.r.t. $t$. It is apparent now that Hamilton's equations (\ref{eq:Hamilton_Equations}) are just the local flow equations of the symplectic Hamiltonian vector field $X_H$ and $\phi^t_H(z_0)$ is the unique trajectory or future development in phase space of the Hamiltonian system with initial data $z_0$. Finally, we define a Hamiltonian system itself as being given by the triple $(\Pi, \omega, X_H)$. We also note that while we have used local coordinates here, all of the above can be stated in a coordinate-independent manner, making it explicitly symplectomorphism-invariant.
 

\subsection{Stability in Classical Mechanics} \label{sec:Stability_CM}
In classical mechanics, given a particular Hamiltonian system $(\Pi, \omega, X_H)$, equilibrium corresponds to a stationary state for variables describing such a system. This means that if a system has initial data $z_0 = z_\star$, and under Hamiltonian evolution one finds a stationary solution $z^a(t) = z^a_\star$, then $z_\star \in \Pi$ is an equilibrium point of the dynamical system. It is clear from the flow equations (\ref{eq:Hamiltonian_Flow}) that the Hamiltonian vector field must have a critical point there, i.e. we have $X_H(z_\star) = 0$. One can now inquire after the nature of the stability of $z_\star$. Such questions are typically concerned with the nature of future developments of initial data `close' to it; i.e. under a `moderate' change in initial data near a critical point, does the trajectory change drastically?

Alternate notions of stability of mechanical systems exist (see for example \cite{Oliveira99}). For example, the Kolmogorov-Arnold-Moser theory \cite{Kolmogorov54, Moser62, Arnold63a, Arnold63b} is an example of a framework within which stability of flows against moderate changes in the Hamiltonian function $H$ itself are dealt with. This sort of a \textit{structural} stability analysis is important in various scenarios \cite{Moser73, Whiteman77, Arnold+06}. Here however we shall purely concern ourselves with stability of future time evolutions against changes in initial data. To qualify the nature of the stability of equilibrium points, we provide below the formal definitions of the notions of non-linear (or Lyapunov), linear and mode (or spectral) stability. There exist other closely related notions of stability, namely of asymptotic or of exponential stability, but we will not go into these here (see \S1 of \cite{Dafermos+16} for a description of these ideas, in the context of general relativity). \\~\\
\noindent
\textbf{Definition:} A critical point $z_\star \in \Pi$ of a Hamiltonian system $(\Pi, \omega, X_H)$ is non-linearly stable (Lyapunov stable) if for every neighborhood $U \subset \Pi$ of $z_\star$ there is a neighborhood $V \subset U$ of $z_\star$ such that for every $z \in V$ the corresponding orbit of the Hamiltonian vector field $\phi^t_H(z)$ remains in $U$ for all $t \geq 0$. Further, if $z_\star$ is not stable, it is unstable.\\
\\
A non-linear stability analysis of a critical point of a dynamical system is, in general, highly non-trivial since for arbitrary Hamiltonian functions, the associated vector field could define a non-linear flow and one is forced to look for solutions to complicated non-linear differential equations. Therefore, the extent of the stability (the size of the region of attraction and the behaviour of transients as they approach the equilibrium, for instance) is typically determined by the non-linearities of the system. However, due to the complexities involved in a non-linear stability analysis, as a preliminary measure, one considers the relatively simpler notions of linear stability.

A linear stability analysis is concerned with studying the behaviour of the future developments of initial data chosen within an \textit{infinitesimal} neighbourhood $\delta U_\star \subset \Pi$ of the equilibrium point $z_\star$, and whether or not they converge to $z_\star$ at late times $t \rightarrow \infty$. For $z \in \delta U_\star$, we now introduce $\xi^a(t) \equiv z^a(t) - z^a_\star$. Then since $|\xi^a(0)| \ll 1$, the Hamiltonian flow equations (\ref{eq:Hamiltonian_Flow}) can be rewritten as,
\begin{equation}
\dot{z}^a_\star + \dot{\xi}^a = X^a_H\left(z_\star + \xi\right) = X^a_H(z_\star) + \left(L_{H\star}\right)^a_{\ b}\xi^b + O(|\xi^a|^2),
\end{equation}
where $L_{H\star}$ is the Jacobian of the symplectic Hamiltonian vector field $X_H$ at $z_\star$ and given as,
\begin{equation}
\left(L_{H\star}\right)^a_{\ b} \equiv \partial_b X^a_H(z_\star) = \partial_b\Omega^{ac}\partial_c H(z_\star) = \Omega^{ac}\partial_c\partial_b H(z_\star).
\end{equation}
From the above, it can be seen that $L_{H\star}$ can also be thought of as being the symplectic Hessian of the Hamiltonian function $H$. $L_{H\star}$ is a constant Hamiltonian matrix, i.e. it satisfies $L_{H\star}^T \Omega + \Omega L_{H\star} = 0$. Then, the equation of motions near a critical point, to leading order in $|\xi^a|$, are given as,
\begin{equation} \label{eq:Linearized_EoM}
\dot{\xi}^a  = \left(L_{H\star}\right)^a_{\ b}\xi^b,
\end{equation}
The above set of equations (\ref{eq:Linearized_EoM}) are called the linearized equations of motion (about a fixed point of the Hamiltonian vector field), and are simply the flow equations of $L_{H\star}$, which is therefore called the linearization of the symplectic Hamiltonian vector field at the equilibrium point (see for example \cite{Bruni+97}). A solution $\xi(t)$ to these flow equations is called the tangent flow and we are now in a position to describe the statement of linear stability of a critical point of a Hamiltonian system.\\~\\
\noindent
\textbf{Definition:} A critical point $z_\star \in \Pi$ of a Hamiltonian system $(\Pi, \omega, X_H)$ is linearly stable if all orbits of the tangent flow are bounded for all forward time.\\

That is, to gauge the linear stability of a dynamical system against perturbations in initial data, we want to know whether the size of arbitrary solutions $\xi$ grows, stays constant, or shrinks as $t \rightarrow \infty$. At simple critical points of the symplectic Hamiltonian vector field $z_\star$, the linearization $L_{H\star}$ is non-singular, and under the assumption of distinct eigenvalues, we can write,
\begin{equation} \label{eq:xi_t}
\xi(t) = \sum_{a=1}^{2n} c_a e^{i\sigma_a t} v_a, 
\end{equation}
where $i \sigma_a$ and $v_a$ are the eigenvalues and eigenvectors of $L_{H\star}$ respectively. We have introduced the additional factor of $i$ to match the usual convention in physics, and our discussion henceforth will be in terms of $\sigma$. It is not difficult to show that the eigenvalues of a Hamiltonian matrix come in $\pm$ pairs \cite{Arnold80}. Consequently, (\ref{eq:xi_t}) has exponentially growing or decaying terms unless all $\sigma$ lie on the real axis. When $L_{H\star}$ is singular, to perform a linear stability analysis, one must naturally look at flows of the first non-singular higher-order symplectic derivative of the Hamiltonian function at the critical point $\partial_{c_1}\partial_{c_2}\cdots\partial_{c_n}\Omega^{ab}\partial_b H(z_\star)$. 

However, studying the forward time boundedness of \textit{every} solution $\xi(t)$ of the linearized equations of motion is a demanding prospect, and a more tractable endeavour is to carefully examine the distribution of the eigenvalues of the Hamiltonian matrix $L_{H\star}$, to find which eigenvectors represent stable and unstable directions in phase space. Now, the weakest statement (though still extremely useful) that can be made regarding the stability of a dynamical system is then as follows.\\~\\
\noindent
\textbf{Definition:} A critical point  $z_\star \in \Pi$ of a Hamiltonian system $(\Pi, \omega, X_H)$ is mode stable (spectrally stable) if all eigenvalues of its corresponding linearization $L_{H\star}$ lie in the left-half plane (or equivalently when all $\sigma$ lie in the upper-half plane).\\
 
Then, eigenvectors corresponding to real $\sigma$ are called normal modes and the eigenvalues $\sigma$ themselves are called normal frequencies. Normal modes are the fundamental oscillatory modes of any conservative system. Similarly, eigenvectors corresponding to $\sigma$ with non-zero imaginary parts are called quasi-normal modes (QNMs) and the corresponding eigenvalues $\sigma$ are called quasi-normal frequencies (QNFs). These are the fundamental oscillatory modes of every dissipative system and the QNF is a complex number with two pieces of information: its real part corresponds to the temporal oscillation and its imaginary part captures the temporal rate of growth or decay. 
Therefore, from the properties of the linearization $L_{H\star}$, one can identify the local stable and unstable manifolds of the critical point, denoted by $E_{\star s}$ and $E_{\star u}$, which are linear subspaces of $T_{z_\star}\Pi$, of dimensions given by the number of stable ($\text{Im}(\sigma) \geq 0$) and unstable eigenvalues respectively, and spanned by the relevant eigenvectors and (in the degenerate case) generalised eigenvectors \cite{Dettmann17}.


We note here a well-known comment regarding the relation between mode and linear stability. Mode stability excludes a particular type of exponentially growing solution; it does not rule out exponential growth in general let alone show that solutions are bounded or decay. The latter would correspond to full linear stability. The precise relation between the two, for autonomous Hamiltonian systems, is that an equilibrium is linearly stable if and only if it is mode stable and all the Jordan blocks of the associated linearization matrix $L_{H\star}$ are one-dimensional (\cite{Abraham_Marsden78, Arnold+06}). 

In summary, when examining the stability of a critical point, the first useful thing to do is to work in the linear approximation. However, for real applications, sometimes just a linear stability analysis can be very misleading and one must therefore pursue a study of the full non-linear stability of equilibria. Thus, nonlinear stability $\Rightarrow$ linear stability $\Rightarrow$ mode stability. Hamiltonian systems that exhibit resonance are classic examples of dynamical systems whose equilibrium configurations are mode stable but linearly unstable (see for example p. 33 of \cite{Marsden_Ratiu99}). An example of a Hamiltonian system that has a critical point that is linearly stable but non-linearly unstable is the famous Cherry Hamiltonian \cite{Cherry26}.

Following \cite{Milnor85}, we now introduce the notions of a dynamical attractor and its basin of attraction; their use will become apparent in \S\ref{sec:Stability_GR} when we discuss the stability of spacetimes in GR, and families of spacetimes. A critical point $z_\star$ of a dynamical system is called an attractor if points in some strictly positive measure neighbourhood $\rho(z_\star) \subset \Pi$, called its realm of attraction, flow to $z_\star$ in forward asymptotic time, i.e. for every $z \in \rho(z_\star)$, $\lim_{t \rightarrow \infty} \phi_H^t(z) \rightarrow z_\star$. If additionally the realm of attraction of such an attractor has the topology of an open set in $\Pi$, then it is typically called a basin of attraction, which we shall denote by $B(z_\star)$. Now, clearly, such an attractor corresponds to a linearly stable critical point, since $B(z_\star)$ contains within it an infinitesimal open neighbourhood of the attractor $\delta U_\star$. Further, the size of the basin of attraction of such an attractor is indicative of the extent of its non-linear (or Lyapunov) stability. If instead $\rho(z_\star)$ is a lower dimensional smooth manifold, then it is generally called the stable manifold of the attractor, akin to $E_{\star s}$ introduced above (note however that $E_{\star s}$ has been defined to be a \textit{local} stable manifold). Such types of attractors were first discussed in \cite{Coddington_Levinson55, Mendelson60, Auslander+64}.

Further, when a dynamical system admits a compact set of critical points in phase space, one can introduce the notion of a compact invariant set $C \subset \Pi$. Such sets $C$ may be regarded as being generalizations of critical points \cite{Auslander+64}. If points in some strictly positive definite measure neighbourhood of $C$, which we will denote by $\rho(C)$ and call the realm of attraction of $C$, flow to $C$, i.e. $C = \bigcup \phi_H^t(z)$ for $z \in \rho(C)$, then $C$ is called a dynamical attractor. Completely analogous criterion for when a compact invariant set is an attractor with a basis of attraction are discussed in \cite{Milnor85}, and the size of its basis of attraction determines the extent of its non-linear stability.

\section{Stability in General Relativity} \label{sec:Stability_GR}
We begin with a brief review of the initial value formulation of general relativity, which is the typical setting of stability analyses in general relativity \cite{Dafermos_Rodnianski13, Dafermos14}. To translate the statements regarding stability into the framework of symplectic geometry, one needs to discuss what the phase space \cite{Ashtekar_Horowitz84}, the symplectic structure \cite{Crnkovic88} and the Hamiltonian \cite{Arnowitt+62} of GR are, thus forming the Hamiltonian system of interest $(\Pi, \omega, X_H)$. Then, by analogy with the analysis for classical mechanics presented earlier, it will become clear how one can think about the usual notions of stability of a given spacetime. We will also discuss the Cauchy problem of the Einstein field equations \cite{Rendall05, Rodnianski06,  Ringstrom09, Christodoulou09} to indicate \textit{when} one can use symplectic geometry to discuss stability. In our discussion here, we shall concern ourselves primarily with spacetimes devoid of matter i.e., the only degrees of freedom of such a system are purely gravitational and the sole dynamical entity is the metric tensor $g$. At the end of this section, we will mention how one could carry this analogy through to include matter.

The usual approach to treating GR as a field theory is based on the covariant Lagrangian formulation and the Einstein-Hilbert action for this system is given as (we have left out the boundary terms),
\begin{equation}
\mathcal{S} = \int \text{d}^4 x \sqrt{-|g|}R,
\end{equation} 
where $|g| \equiv \text{det}(g_{\mu\nu})$ is the determinant of the metric tensor, $R$ is the associated Ricci scalar and we have used Geometrized units, $16\pi G \! = c \! = \! 1$. With the introduction of the associated Christoffel connections $\Gamma^\mu_{\ \rho\sigma} = \frac{1}{2}g^{\mu\alpha}\left(2 g_{\alpha(\rho,\sigma)} - g_{\rho\sigma,\alpha}\right)$, the Riemann tensor $R^{\rho}_{\ \sigma\mu\nu} = 2\left(\Gamma^\rho_{\ \alpha[\mu}\Gamma^\alpha_{\ \nu]\sigma} - \Gamma^\rho_{\ \sigma[\mu,\nu]}\right)$ and the Ricci tensor $R_{\mu\nu} = R^{\rho}_{\ \mu\rho\nu}$, the Ricci scalar is given as $R = g^{\mu\nu}R_{\mu\nu}$. Then the governing Euler-Lagrange equations of motion are,
\begin{equation} \label{eq:VEEs}
R_{\mu\nu}(g) = 0,
\end{equation}
where in the above square brackets mean antisymmetrization in the outermost indices $[\mu \cdots \nu] = \frac{1}{2}(\mu \cdots \nu - \nu \cdots \mu)$ and round brackets mean symmetrization similarly defined, $(\mu \dots \nu) = \frac{1}{2}(\mu \cdots \nu + \nu \cdots \mu)$. The above set of equations are called the vacuum Einstein field equations and metrics $g$ that satisfy these equations are called the vacuum solutions of GR. We will now first discuss what forms a valid initial data set for these dynamical equations.

A triple ($\Sigma, q, K$) with $\Sigma$ a smooth orientable 3-manifold, $q$ a Riemannian metric and $K$ a symmetric 2-tensor field, both on $\Sigma$, forms a valid initial data set for the vacuum Einstein field equations (\ref{eq:VEEs}) if the geometry initial data fields $d_g \equiv (q, K)$ satisfy certain constraint equations. Given such an initial data set, one expects to generate a 4-dimensional spacetime $(\mathcal{M}, g)$ and a one-parameter family of embeddings $\theta: \Sigma \times \mathbb{R} \rightarrow \mathcal{M}$ 
such that $g$ satisfies the EFEs (\ref{eq:VEEs}), $\theta(\Sigma)$ forms a Cauchy hypersurface in $\mathcal{M}$, and the fields $q$ and  $K$ are the first and second fundamentals respectively of $\Sigma$ in $(\mathcal{M}, g)$  (see for example p. 226 of \cite{Hawking_Ellis73}). 
The constraint equations mentioned above are simply the mathematical consequences of the desire that $\Sigma$ `fit properly' into $\mathcal{M}$, i.e. $d_g$ satisfies what are called the Gauss and Codazzi-Mainardi equations, which govern the embeddings of hypersurfaces into manifolds. In the context of GR, these are usually called the Einstein Hamiltonian and momentum constraint equations and are given respectively as (see for example \cite{Blau18}),
\begin{align} \label{eq:Constraints}
^{(3)}\!R - K_{ij}K^{ij} + \left(K^i_{\ i}\right)^2 = 0, \\
\nabla^{j}K_{ij} - \nabla_{i}K^j_{\ j} = 0,
\end{align}
where $^{(3)}\!R$ is the Ricci scalar and $\nabla$ the covariant derivative associated with $q$. For insight into the structure of the constraint differential equations we direct the reader towards \cite{Rendall05, Rodnianski06, Ringstrom09}. This property that initial data fields cannot be freely specified but must satisfy certain constraint equations is not characteristic to GR and, for example, is also a feature of Maxwell's equations for electromagnetism, where the constraint equation is the Gauss law \cite{Henneaux_Teitelboim92}.

Technically, $(\mathcal{M}, \theta, g)$ is called a development of $(\Sigma, d_g)$ and the evolution or Cauchy problem in GR refers to the construction of the former from the latter. The Cauchy problem of a given field theory is well-posed if for any valid choice of initial data, there exists a solution which is consistent with that data, and the map from the space of initial data to solutions is continuous \cite{Rodnianski06}. 
Choquet-Bruhat \cite{Foures-Bruhat52, Bruhat62} showed that the Cauchy problem in GR is indeed well-posed. Further, Choquet-Bruhat \& Geroch \cite{Choquet-Bruhat_Geroch69} showed that each such initial data set has a unique maximal future development $(\mathcal{M}, g)$, namely a development which extends every other development of the same initial data set%
\footnote{Another development $(\mathcal{M}^\prime, \theta^\prime, g^\prime)$ of $(\Sigma, d_g)$ is called an extension of $\mathcal{M}$ if there is a diffeomorphism $\alpha$ of $\mathcal{M}$ into $\mathcal{M}^\prime$ such that $\theta^\prime(\Sigma) = (\alpha \circ \theta)(\Sigma)$ and $\alpha_* g^\prime = g$. In particular, a maximal extension is an extension of any development of $(\Sigma, d_g)$ \cite{Hawking_Ellis73}.}. %
Geroch \cite{Geroch70} subsequently showed that for any such development $(\mathcal{M}, g)$, the manifold $\mathcal{M}$ is diffeomorphic to $\Sigma \times \mathbb{R}$, and such spacetimes are called globally hyperbolic spacetimes. 
 These fundamental results are critical to argue the existence of solutions in GR and to set up the Hamiltonian formulation of GR.

In this context, the stability of a spacetime $(\mathcal{M}, g)$ is understood as follows. First one finds the initial data $d_g$ whose evolution under the Einstein field equations yields $(\mathcal{M}, g)$. Then, in the space of all allowed initial data, one considers neighbourhoods around $d_g$ and checks whether their future developments yield metrics $g^\prime$ such that ``$g^\prime \approx g$,'' in some sense. Depending on whether or not the neighbourhoods under consideration are infinitesimal or not, one is conducting then either a linear or non-linear stability analysis respectively. For example, the Kerr family of spacetimes ($\mathcal{M}, g_{M, a}$) is a 2-parameter family of solutions of the Einstein field equations and contains the 1-parameter family of Schwarzschild solutions denoted by $g_{M, 0}$. Within this family is also the Minkowski metric for which the metric can be represented as $g_{0,0}$. Therefore, if a particular member $g_{M, a}$ were generated by some initial data $d_{M, a}$ and one considers another initial data within an infinitesimal neighbourhood of $d_{M, a}$, which let us denote by $d_{M + \delta M, a + \delta a}$, then if the future development of the latter $g_{M + \delta M, a + \delta a}$ is such that $|g_{M + \delta M, a + \delta a} - g_{M, a}| \ll 1$, then one can say that the spacetime $g_{M, a}$ is linearly stable. If such a statement holds for all $M, a$, then one can argue that the family of Kerr spacetimes is linearly stable. To find the full extent of stability of a particular solution $g_{M, a}$, one must conduct a non-linear stability analysis which requires one examine to the future developments of initial data chosen within arbitrary, non-infinitesimal neighbourhoods of $d_{M, a}$. These issues of stability are extremely important, both from theoretical and astrophysical standpoints, and difficult to examine and thus far, significant progress has been made and the non-linear stability of the Minkowski spacetime \cite{Christodoulou_Klainerman93} and the linear stability of the Schwarzschild spacetime \cite{Dafermos+16} has been established, and it has also been shown that the Kerr family of spacetimes are mode stable \cite{Whiting89}. For more detailed and excellent discussions on stability see \cite{Dafermos14}.

Now, to restate all of these notions in the language of symplectic geometry, we need to define the conjugate momentum to $q$ and identify the phase space of GR. For this, one requires an explicit notion of time, which has the implication that one can only consider spacetimes that are topologically of the form $\Sigma \times \mathbb{R}$, where $\Sigma$ is a 3-dimensional manifold of arbitrary, fixed topology and $\mathbb{R}$ is time. Due to the discussion presented above, this restriction is typically not considered a strong limitation. However, it is useful to remember that not all solutions of the Einstein field equations are globally hyperbolic. For example, the maximally extended Reissner-Nordstrom solution, representing the spacetime for a spherically symmetric charged particle, has a Cauchy horizon i.e., there is a region from which there exist past directed causal curves that do not pass through any candidate Cauchy surface. Generally such solutions are discarded as being physically unrealistic since a desirable quality of physical theories is that they be deterministic and that there exist a one-to-one map between the initial state of motion and its trajectory (`a continuous dependence of the evolution on initial data'). However, it is still unclear whether only globally hyperbolic spacetimes should be considered in GR \cite{Witten19}. Here however we will restrict ourselves to a discussion of the stability of only globally hyperbolic spacetimes.

Given a particular initial data set $(q, K)$ that belongs to the space of valid initial data, one can define the corresponding element $(q, p)$ of the phase space of general relativity $\Pi$ via (see for example \S 20 of \cite{Blau18}),
\begin{equation} \label{eq:p_def}
p^{ij} = \sqrt{|q|}(K^{ij} - q^{ij}K^k_{\ k}),
\end{equation}
where $|q| \equiv \text{det}(q_{ij})$ and $p$, the momentum conjugate to $q$, is a symmetric 2-tensor. Then, we can write the Hamiltonian density function as \cite{Blau18},
\begin{equation}
\mathcal{H}_{\text{ADM}} = \alpha\left[- \sqrt{|q|}~ ^{(3)}\!R + \frac{1}{\sqrt{|q|}}\left(p^{ij}p_{ij} - \frac{1}{2}|p|^2\right)\right] - 2\beta_j\nabla_ip^{ij},
\end{equation}
where $\alpha$ and $\beta$ are the lapse function and the shift vector respectively, as usual, and $|p| = \text{det}(p^{ij})$. The ADM Hamiltonian is given as $H_{\text{ADM}}  = \int_\Sigma \mathcal{H}_{\text{ADM}}$. With little effort, it is evident that the variation of $H_{\text{ADM}} $ w.r.t $\alpha, \beta$ simply gives the constraint equations (\ref{eq:Constraints}), rewritten in terms of $q, p$. The variation w.r.t the dynamical degrees of freedom%
\footnote{Note that the lapse function $\alpha$ and shift vector $\beta$ are not dynamical because they describe how coordinates move in time from one hypersurface to the next and have to be fixed by four gauge conditions (see for example \S3.9 of \cite{Straumann13}). One simple choice corresponds to the Gaussian normal coordinates, for example, where one sets $\alpha = 1, \beta = 0$.} %
$q, p$ now gives the Hamilton equations for GR,
\begin{align}
\dot{q}_{ij} =& \frac{2\alpha}{\sqrt{|q|}}\left(p_{ij} - \frac{1}{2}q_{ij}|p|\right) + \nabla_{(i}\beta_{j)}, \\
\dot{p}^{ij} =& -\alpha\sqrt{|q|}\left(^{(3)}\!R^{ij} - \frac{1}{2}~\!^{(3)}\!R q^{ij}\right) + \frac{\alpha q^{ij}}{2 \sqrt{|q|}}\left(p_{mn}p^{mn} - \frac{1}{2}|p|^2\right) - \frac{2\alpha}{\sqrt{|q|}}\left(p^{ik}p_{k}^{\ j} - \frac{1}{2}|p|p^{ij}\right) \\ 
&+ \sqrt{|q|}(\nabla^i\nabla^j\alpha - q^{ij}\nabla^{i}\nabla_{i}\alpha) + \sqrt{|q|}\nabla_{k}\left(\frac{\beta^k p^{ij}}{\sqrt{|q|}}\right) - 2 p^{k(i}\nabla_k \beta^{j)}. \nonumber
\end{align}
The above equations are just the flow equations of the symplectic Hamiltonian vector field obtained from the ADM Hamiltonian function $H_{\text{ADM}}$. However, before one can study the stability of its critical points, one must worry about gauge degeneracies and the construction of the reduced phase space by forming the quotient space of the constrained phase space with the gauge orbits. Then on reduced phase space, one would be able to successfully draw a formal analogy between the notions of stability in classical mechanics to those in general relativity. This will be attempted elsewhere and here we only present partial results as motivation for the extended, deeper study.

It is clear that the Minkowski metric $g_{0,0}$, a stationary solution of the Einstein field equations, is a critical point in the phase space of the ADM Hamiltonian dynamical system. Similarly, it can be seen then that the Schwarzschild one-parameter family of solutions $g_{M,0}$ is a compact invariant set and the Kerr two-parameter family of metrics $g_{M,a}$ forms an even larger compact invariant set in phase space. Now, the results of Christodoulou \& Klainerman \cite{Christodoulou_Klainerman93} imply that the Minkowski solution is a dynamical (critical point) attractor in phase space and the results of Dafermos, Holzegel \& Rodnianski \cite{Dafermos+16} indicate that the Schwarzschild family of solutions is also an (compact invariant set) attractor in phase space. The basin of attraction of the Schwarzschild attractor remains to be completely characterized, and we make a restricted attempt to address this in \S\ref{sec:OSD_Non-Linear}. Now, we can express one of the most important aims of stability studies in general relativity, with significant implications for observational astrophysics, as being to show that the full Kerr family of solutions is an attractor, and a complete characterization of its basin of attraction will conclude a non-linear stability analysis of the Kerr family of spacetimes.

When matter is present, the full classical action of this system is given by the Einstein--Hilbert Lagrangian plus a piece describing the matter fields $\phi_i$ appearing in the theory, and extremizing this action with respect to the metric tensor $g_{\mu\nu}$ yields,
\begin{equation}
\mathbb{G}_{\alpha\beta} \equiv R_{\alpha\beta}(g) - \frac{1}{2}g_{\alpha\beta}R(g) = \frac{1}{2}T_{\alpha\beta},
\end{equation}
where we have introduced the Einstein tensor $\mathbb{G}_{\alpha\beta}$ and the matter energy-momentum tensor $T_{\alpha\beta}$. Naturally, one needs to include the coupled matter equations of motion arising from the extremization of the action w.r.t. the matter fields $\phi_i$ to obtain the full set of equations that govern the dynamics of this system. In general, this set of Einstein plus matter equations of motion do not form a closed set of partial differential equations and one is required to introduce a constitutive relation determining the energy-momentum tensor $T_{\mu\nu}$ from the metric and the matter fields $g$ and $\phi_i$. These equations and relations can be constructed from the appropriate classical field theory describing the matter model of interest, scalar or electromagnetic fields or hydrodynamic fluids etc. (see for example \cite{Hawking_Ellis73, Rodnianski06}). The initial data then for the associated Cauchy problem will is given by $d = d_g \cup d_m$, where $d_m$ denotes the initial data for the matter sector.

In the following section, we demonstrate a typical non-linear stability analysis in general relativity, in the setting of a simplistic spherically symmetric collapse model. We will take our matter model to be that of a fluid with vanishing pressure since for such systems, the mass contained within a shell of arbitrary comoving radius is conserved throughout the collapse, leading to substantial simplifications in the dynamical equations, making a purely analytical approach tractable. 

\section{Stability in Gravitational Collapse: A Demonstration} \label{sec:Stab_GC}
Armed now with an understanding of the formal notions of stability of dynamical systems, in this section we will argue that the study of the genericity of formation of particular black hole or naked singularity stationary solutions, from the gravitational collapse of regular matter, is essentially a characterization of their realms of attraction. Our approach here will be to revert back to the usual initial value formulation of GR, to provide a demonstration of this statement by analysing the extent of the basin of attraction of the Schwarzschild attractor solution. This can equivalently be thought of as a study of the divergence of the flow of the symplectic Hamiltonian vector field of GR (coupled to matter fields) near the initial data of the collapse process that terminates in a Schwarzschild black hole (a non-linear stability analysis).

The Oppenheimer-Snyder-Datt (OSD \cite{Datt38, Oppenheimer_Snyder39}) solution of the Einstein field equations evolves a regular, homogeneous, spherically symmetric ball of dust (pressureless fluid) to a Schwarzschild black hole. We will see that the prescription of initial data for this process corresponds to specifying the initial density,
\begin{equation}
\rho(t=0, r) = \rho_0 = \text{const.},
\end{equation}
and the initial binding energy profile of the dust cloud, characterised by $f(r)$. In \S\ref{sec:OSD_Non-Linear}, we perform a preliminary non-linear stability analysis of the marginally bound ($f = 0$) OSD collapse processes against a \textit{specific} class of deformations from homogeneity in the initial density profile. What we mean by this is that we will consider initial density profiles $\rho(0,r)$ of the form,
\begin{equation} \label{eq:Inhomogeneity}
\rho(0,r) = \rho_0 - \rho_2 r^2,\ \ \rho_2 \geq 0,
\end{equation}
(for $\rho_2 = 0$, we recover the OSD collapse process) and study the local and global visibility of the eventual singularity that forms as a result of continual collapse. This will give us the size of the region in the (restricted) space of initial data (i.e., the $\rho_0$-$\rho_2$ parameter space), around $(\rho_0, 0)$, whose evolutions under Einstein field equations result in black holes. If the size of this region is an infinitesimal open set around $(\rho_0, 0)$, we will conclude that the OSD collapse process is linearly stable against changes in initial data. However, if this region is larger, then this will provide us with a measure of the extent to which the OSD collapse process is non-linearly stable against changes in initial data.

Now, since it is known that the marginally bound OSD collapse processes sit as a one-parameter ($\rho_0$) subclass of the marginally-bound Lema{\^i}tre-Tolman-Bondi (LTB  \cite{Lemaitre33, Tolman34, Bondi47}) solutions, which have an entire square integrable-function's worth of freedom (corresponding to the initial density profile of the matter $\rho(0, r)$) that can be freely prescribed, the evolutions of the initial data corresponding to the inhomogeneous initial density profiles given in (\ref{eq:Inhomogeneity}) are already known. We avoid the term proportional to $r$ to avoid density cusps at the centre of the collapsing cloud (see \S\ref{sec:Initial_Data} below). Now, following \cite{Joshi_Dwivedi93}, we analyse the structure of the singularity that forms in the class of marginally bound LTB collapse models with initial density profiles given as (\ref{eq:Inhomogeneity}). A more exhaustive non-linear stability analysis of the formation of a Schwarzschild black hole as an end-state of the Oppenheimer-Snyder-Datt (OSD) collapse process incorporating recent results \cite{Joshi_Malafarina11a, Joshi_Malafarina11, Joshi_Malafarina15, Szekeres75} will be reported elsewhere.

\subsection{Dynamics of Dust Collapse} \label{sec:EEs_Vanishing_Pressure}
The spacetime geometry associated with a spherically symmetric collapsing cloud of matter is described by the (interior) metric,
\begin{equation} \label{eq:GCM}
ds^2 = -e^{2\nu(t,r)}~dt^2 + \frac{R^{\prime 2}(t,r)}{1 + f(t,r)}~dr^2 + R(t,r)^2~d\Omega^2,
\end{equation}
where in the above $(t,r,\theta, \phi)$ are Langrangian coordinates, comoving with the matter field, i.e. in these coordinates the matter four-velocity is given as $u^\alpha = e^{-\nu}\delta^\alpha_{\ t}.$ The range of the radial coordinate is $0 \leq r \leq r_{\text{B}}$, where $r_{\text{B}}$ is the boundary of the matter cloud and we have written $g_{rr}$ in this form, anticipating convenience. The $^\prime$ denotes a derivative w.r.t. $r$ and
\begin{equation}
d\Omega^2 = d\theta^2 + \sin^2\theta~d\phi^2
\end{equation}
is the standard metric on a unit two-sphere. The metric function $\nu$ is related to the redshift, $R$ is the proper radius of a shell of collapsing matter present at a comoving radius $r$ and time $t$, and $f$ characterizes the binding energy profile of the collapsing cloud \cite{Joshi93}. In anticipation of its immediate use, we now introduce the Misner-Sharp mass function $F(t,r)$ which measures the amount of mass contained within a shell of comoving radius $r$ at a time $t$ is given as \cite{Misner_Sharp64, Cahill_McVittie69}, 
\begin{equation} \label{eq:MSMass}
F(t,r) \equiv R\left(1 - g^{\alpha\beta}\partial_\alpha R~\partial_\beta R\right)  = R\left(e^{-2\nu}\dot{R}^2 - f(t,r)\right),
\end{equation}
where $g$ is the metric tensor obtained from (\ref{eq:GCM}). Since this metric only describes a portion of the spacetime ($0 \leq r \leq r_{\text{B}}$), if one wants to consider the collapse of matter that has compact support on the initial spacelike hypersurface $t=0$, to complete the spacetime one must match this interior collapsing metric at the boundary with an appropriate exterior metric, which via Birkhoff's theorem \cite{Birkhoff23}, must necessarily be the Schwarzschild metric. 

The interior collapsing metric $g$ contains an apparent horizon, which is the marginally trapped surface, if the radial null expansion scalar, defined as \cite{Poisson04}
\begin{equation}
\theta(t, r) \equiv g^{\alpha\beta}\partial_\alpha R~\partial_\beta R
\end{equation}
vanishes. Therefore, it is seen from (\ref{eq:MSMass}) that the apparent horizon curve $t_{\text{AH}}(r)$, which tracks the location of the apparent horizon during the evolution of the collapse can be found from,
\begin{equation} \label{eq:tAH}
F(t_{\text{AH}}(r), r) = R(t_{\text{AH}}(r), r).
\end{equation}
Now, if we consider the spherically symmetric collapse of a fluid with vanishing pressure in this choice of comoving coordinates, we can write the associated matter stress-energy tensor as,
\begin{equation} \label{eq:TmunuTransverse}
T^\mu_{\ \nu} = \text{diag}(-\rho, 0, 0, 0).
\end{equation}
The choice to consider fluids with vanishing pressure $p_r = p_\theta = 0$ greatly simplifies the collapse evolution (see for example \S4.2 of \cite{Joshi_Malafarina11}). Firstly, $F = F(r)$ and $f = f(r)$ become time independent and are therefore completely set by their initial values; $F, f$ are no longer dynamical functions. Further, $\nu$ does not depend on $r$ and $\nu = \nu(t)$ i.e., by rescaling time for the interior metric, we can set $\nu = 0$. Then the governing EFEs, $\mathbb{G}^\mu_{\ \nu} = T^\mu_{\ \nu}$, for the evolution of such a fluid are given as,
\begin{eqnarray}  
\rho &=& \frac{F^\prime}{R^2 R^\prime}, 
\label{eq:rhoEE} \\
\dot{R} &=& -\sqrt{\frac{F}{R} + f},
\label{eq:dotR}
\end{eqnarray}
where in the above we have rewritten (\ref{eq:MSMass}) as (\ref{eq:dotR}) and have chosen the negative root since we are interested here in collapsing solutions. 

Also, throughout the collapse process we shall require that the weak energy condition is satisfied everywhere i.e. $T_{\mu\nu}v^\mu v^\nu \geq 0$ for all non-spacelike vectors $v^\mu$. This implies that the energy density is everywhere positive $\rho \geq 0$, including near $r=0$. Singularities are points of spacetime where the usual differentiability and manifold structures break down. They are characterized by divergences in the matter energy density or curvature invariants constructed from the Riemann curvature tensor, like the Kretschmann scalar $\kappa \equiv R_{\mu\nu\rho\sigma}R^{\mu\nu\rho\sigma}$. As can be seen from (\ref{eq:rhoEE}), the energy density diverges when $R=0$ or when $R^\prime = 0$. The latter condition corresponds to the collision of different radial shells of matter, which cause what are known as `shell-crossing singularities.' These types of singularities are weak singularities and are removable by a suitable change of coordinates  (see for example \S6.8 of \cite{Joshi93} for further discussion; also see \cite{Joshi_Malafarina11}). Therefore, we shall also require that $R$ satisfy $R^\prime \neq 0$. Further, for the weak energy condition to hold on the initial epoch $\rho(0,r) \geq 0$ from which the collapse begins, we require that $F^\prime \geq 0$ from (\ref{eq:Fr}). Now, for collapse processes of interest here we have $F = F(r)$ which means that for the energy condition to hold at all times $\rho(t, r) \geq 0$, we require specifically from (\ref{eq:rhoEE}) that $R^\prime > 0$. Finally, $R(t_{\text{s}}(r), r) = 0$ are genuine spacetime singularities, also called shell-focussing singularities, and $t_{\text{s}}(r)$ is called the singularity curve i.e. it is the time at which the shell at comoving radius $r$ reaches the singularity. Therefore the coordinate time runs from $-\infty < t < t_{\text{s}}(r)$. 

\subsection{Initial Data} \label{sec:Initial_Data}
We now discuss how one sets valid initial data, $d = \{R(0,r), f(r), \dot{R}(0,r), \rho(0, r), F(r), \dot{\rho}(0,r)\}$. We can partition this set of initial data heuristically into geometry $d_g = \{R(0,r), f(r), \dot{R}(0,r)\}$ and matter $d_m = \{\rho(0, r), F(r), \dot{\rho}(0,r)\}$ initial data. By valid initial data, we mean that the set of functions listed in $d$ must be chosen such that they respect the Hamiltonian and momentum constraints, are smooth and are such that no singularity or apparent horizon is present on the Cauchy surface, $t=0$. 

First we shall inquire after the number of independent initial data functions in this collapse model. If, without loss of generality, we choose the initial scaling as $R(0, r) \! = \! r$, then prescribing the initial density profile $\rho(0,r)$ at the initial epoch fixes the matter profile of the cloud $F(r)$ from (\ref{eq:rhoEE}) as,
\begin{equation} \label{eq:Fr}
F(r) = \int_{0}^{r}\rho(0, \tilde{r})\tilde{r}^2\text{d}\tilde{r}.
\end{equation}
Further, picking $f(r)$ fixes $\dot{R}(0,r)$ from (\ref{eq:MSMass}) and $\dot{\rho}(0,r)$ is set then from (\ref{eq:rhoEE}). In summary, one is free only to pick three functions independently on the Cauchy surface $t=0$, which here will be $R(0,r), f(r), \rho(0, r)$. The remaining initial data $\{\dot{R}(0,r), F(r), \dot{\rho}(0,r)\}$ are then fixed from the constraint equations.

Now we move to a discussion on the smoothness of initial data. The requirement that $\rho(0, r)$ be smooth implies that $F$ is atleast $O(r^3)$ near $r=0$. Also, since realistically the centre of the collapsing cloud has non-zero density, i.e. $\rho(0,0) > 0$, we have also $F^{\prime\prime\prime}(0) > 0$. Moreover, for $\dot{R}(0, r)$ to be regular, 
we shall require that 
\begin{equation}
\pi(r) \equiv - \frac{r f(r)}{F(r)} 
\end{equation}
also be regular throughout the dust cloud. Specifically, for $\pi(r)$ to be regular at $r=0$, we require $f$ to be $O(r^2)$ near $r=0$. If we choose the usual scaling $R(0,r) = r$, then the remaining freely specifiable initial data ($f(r), \rho(0,r)$) will be assumed to be atleast square integrable. We will also require that there be no cusps at the centre of the cloud and so, we will impose the restriction that $f(r), \rho(0,r)$ not have terms that are odd powers of $r$ near $r=0$ \cite{Joshi_Singh95}. Further, for no trapped surfaces to exist on the Cauchy surface $t=0$, we require $F(r)/R(0,r) < 1$. Finally, the requirement that there also be no singularity on the Cauchy surface $t=0$ must be discussed on a case by case basis.

\subsection{Non-linear Stability of the Oppenheimer-Snyder-Datt Collapse Process} \label{sec:OSD_Non-Linear}
Since the governing equations of motion (\ref{eq:rhoEE}, \ref{eq:dotR}) are closed, they evolve valid initial data uniquely and depending on the specific choice of initial data, the singularity may or may not be covered entirely from an asymptotic observer by a horizon, corresponding to the formation of a black hole or a globally naked singularity respectively. Here, for simplicity, we shall restrict ourselves to the class of marginally bound LTB models ($f = 0$). For this class of collapse models, we can immediately integrate (\ref{eq:dotR}) to obtain the scale factor $R$ analytically as,
\begin{equation}
R(t,r) = r\left(1 - \frac{3}{2}\sqrt{\frac{F}{r^3}}t\right)^{2/3}.
\end{equation}
The energy density $\rho(t,r)$ is then given as,
\begin{equation}
\rho(t,r) =  \frac{F^\prime}{r^2\left(1 - \frac{3}{2}\sqrt{\frac{F}{r^3}}t\right)\left(1 - \frac{r F^\prime}{2 F}\sqrt{\frac{F}{r^3}}t\right)} .
\end{equation}
The above two equations completely specify the marginally bound LTB collapse models. The singularity $t_{\text{s}}(r)$ and the apparent horizon $t_{\text{AH}}(r)$ curves are obtained from the conditions $R(t_{\text{s}}(r),r) = 0$ and $R(t_{\text{AH}}(r),r) = F(r)$ respectively as,
\begin{equation}
t_{\text{s}}(r) = \frac{2}{3}\sqrt{\frac{r^3}{F}},\ \ t_{\text{AH}}(r) = t_{\text{s}}(r)\left[1 - \left(\frac{F}{r}\right)^{3/2}\right].
\end{equation}
Since we have already seen that $F$ is atleast $O(r^3)$ and always non-negative, it is clear from the above equation that $t_{\text{AH}}(r) < t_{\text{s}}(r)$ for all $0 < r$. Therefore, outgoing null geodesics emitted from events $(t_{\text{s}}(r), r)$ for $r \neq 0$ are all trapped. Now, to determine the causal structure of a particular model, namely whether it represents a black hole or a globally visible naked singularity, it is necessary to examine families of radial null geodesics emerging from the event $(t_{\text{s}}(0),0)$. Our analysis closely follows the procedure outlined in \cite{Joshi_Dwivedi93}. It is evident from (\ref{eq:GCM}) that along future-directed radial null geodesics, we have 
\begin{equation}
\frac{dt}{dr} = R^\prime.
\end{equation}
More importantly, since we are concerned with outgoing null geodesics that are emitted from $(t_{\text{s}}(0), 0)$, we should check whether there exist null geodesics along which $dR/dr > 0$ at $(t_{\text{s}}(0), 0)$, corresponding to a positive future null expansion. Further, since $R$ vanishes as $t\rightarrow t_{\text{s}}(0), r \rightarrow 0$, if there exists such a geodesic, then one can find a positive constant $\alpha$ such that along it, near $(t_{\text{s}}(0), 0)$, we can write $R \sim r^\alpha$. $\alpha$ is initial data dependent and, if it exists, can be found via the procedure outlined below.

Now, to analyse $dR/dr$ it is useful to change variables from $r$ to $u=r^{\alpha}$ so that along null geodesics we can write,
\begin{equation} \label{eq:dRdu}
\frac{dR}{du} = \frac{1}{\alpha r^{\alpha-1}}\left[R^{\prime} + \dot{R}\frac{dt}{dr}\right] = \left[1 - \sqrt{{\Lambda\over X}}\right]\frac{R^{\prime}}{\alpha r^{\alpha-1}} \equiv \left[1 - \sqrt{{\Lambda\over X}}\right]\frac{H(X, u)}{\alpha},
\end{equation}
where we have introduced $\Lambda(u) = F(u)/u$ and $X(R, u) = R/u$. Further, $H(X,u)$ defined as above can be written out as,
\begin{equation} \label{eq:HXu}
H(X,u) = \frac{u^{\frac{3 - 3\alpha}{2\alpha}}}{\sqrt{X}}\left(1 - \frac{\eta}{3}\right) + \frac{\eta X}{3},
\end{equation}
where $\eta(u) = u^{1/\alpha} F^\prime/F$. Note that the event $(t, r) = (t_{\text{s}}(0), 0)$ is now at $(R, u) = (0, 0)$. The above differential equation (\ref{eq:dRdu}) has a singular point at $(X, u) = (0, 0)$ and if there exist null geodesics that meet this singularity, we can write along them,
\begin{equation}
\lim_{R \rightarrow 0, r \rightarrow 0} \frac{R}{u} = \lim_{R \rightarrow 0, r \rightarrow 0} \frac{dR}{du} = X_{0}.
\end{equation}
and we are assured that they are outgoing if $X_0 > 0$. Therefore, the necessary and sufficient condition for the singularity to be visible in dust collapse is that $X_0$, which can be found as the root of the algebraic equation,
\begin{equation}  \label{eq:VX}
X = \left[1 - \sqrt{\frac{\Lambda_0}{X}}\right]\frac{H(X, 0)}{\alpha},
\end{equation}
exists and is positive. Here we have introduced $\Lambda_0 = \lim_{u \rightarrow 0}\Lambda(u)$. The parameter $\alpha$ is chosen, when possible, such that $dR/du$ is well defined along such null geodesics, i.e. $H(X, 0)$ is well defined. 

The existence of a real positive root of equation (\ref{eq:VX}) ensures that a family of outgoing null geodesics terminates at the singularity in the past and guarantees that the singularity is naked; when no real positive roots exist, the singularity is space-like and the spacetime contains an OSD-like black hole. Even when a real positive root $X_0$ of (\ref{eq:VX}) exists, whether or not the singularity is globally visible (visible to asymptotic observers) depends on the initial density profile (or equivalently the mass function $F$), as we discuss below. A singularity is globally visible {\it if and only if} there exist families of outgoing null geodesics that emanate from the singularity and have a positive future null expansion. This condition is given as (see \S III.C of \cite{Joshi_Dwivedi93}),
\begin{equation} \label{eq:Visibility_Condition}
\eta \Lambda < \alpha X_0. 
\end{equation}
The above condition must be satisfied along null geodesics that are outgoing from the the singularity throughout the matter cloud i.e., for $0 \leq r \leq r_{\text{b}}$. Outgoing null geodesics emanating from the singularity that satisfy this condition continue to remain outside the apparent horizon as they move into the future till they reach the boundary of the dust cloud. These trajectories then reach future null infinity in the exterior Schwarzschild region.\\

\begin{figure}
\includegraphics[scale=1]{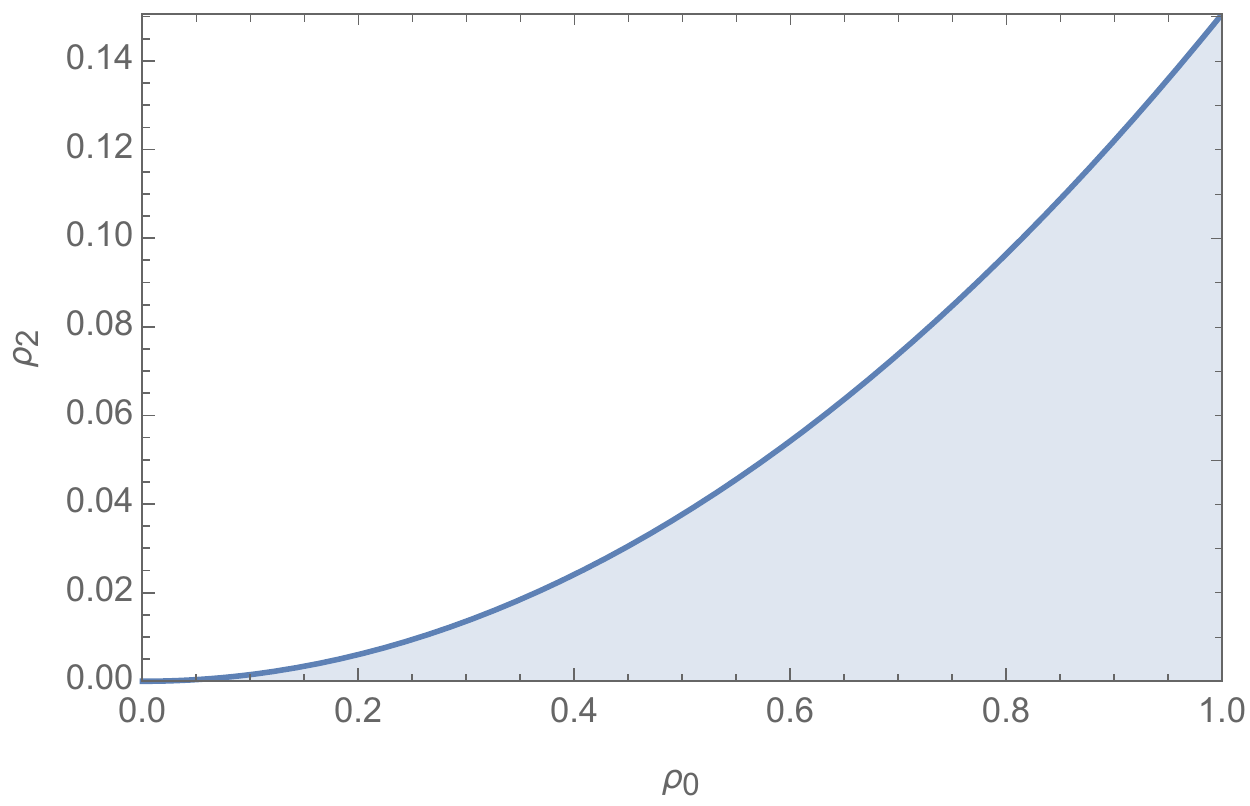}
\caption{The space of initial data of the marginally bound Lema{\^i}tre-Tolman-Bondi dust collapse models with a quadratic initial density profile $\rho(0,r) = \rho_0 - \rho_2 r^2$ ($\rho_2 \geq 0$) is parametrized by $(\rho_0, \rho_2)$. Here we indicate by the shaded region the set of initial data that develop black holes. The complementary region, in white, corresponds to the set of initial data for which the spacetime contains a globally visible naked singularity. Although here we have only displayed the region $0 \leq \rho_0 \leq 1$ of the parameter space, it is evident from (\ref{eq:GlobalVisibility_rho2_rho0}) that this partitioning is representative of the entire parameter space $\rho_0-\rho_2$. Also, since we require that the energy condition $\rho \geq 0$ be satisfied throughout the cloud, its maximum radius cannot exceed $r_{\text{b, max}} = \sqrt{\rho_0/\rho_2}$.}
 \label{fig:Global_Visibility_Figure}
\end{figure}

\noindent
Now, for the choice of initial data we are interested in, the initial density profile is given as, 
\begin{equation}
\rho(0, r) = \rho_0 - \rho_2 r^2,\ \ \text{with}\ \ \rho_2 \geq 0.
\end{equation}
Then we can expand $H(X, u)$ around $u=0$ to leading order for each term in (\ref{eq:HXu}) to obtain,
\begin{equation}
H(X, 0) = \frac{1}{\sqrt{X}}\frac{2\rho_2}{5\rho_0}u^{\frac{7 - 3\alpha}{2\alpha}} + X.
\end{equation}
It is evident that for $H(X, 0)$ to be well defined, we must set $\alpha \leq 7/3$. However, we obtain a real root $X_0$ of the `root equation' (\ref{eq:VX}) only for $\alpha = 7/3$,  which is now given as,
\begin{equation}
\frac{7}{3}X = \frac{1}{\sqrt{X}}\frac{2\rho_2}{5\rho_0} + X.
\end{equation}
That is, $X_0^{3/2} = 3\rho_2/10\rho_0$. Therefore, with the slightest deviation from homogeneity ($\rho_2 > 0$), the LTB singularity becomes locally naked. Further, the condition for global visibility (\ref{eq:Visibility_Condition}) can be written as,
\begin{equation}
\left(\rho_{0} - \rho_{2}r^2\right)r^{2/3} < \frac{7}{3}\left(\frac{3}{10}\frac{\rho_{2}}{\rho_{0}}\right)^{2/3}.
\end{equation}
This inequality must hold for the largest value of the expression on the left, which is attained at $r = \sqrt{\rho_{0}/4\rho_{2}}$. Then we can write,
\begin{equation}
\left(\rho_{0} - \rho_{2}\frac{\rho_{0}}{4\rho_{2}}\right)\left(\frac{\rho_{0}}{4\rho_{2}}\right)^{1/3} < \frac{7}{3}\left(\frac{3}{10}\frac{\rho_{2}}{\rho_{0}}\right)^{2/3},
\end{equation}
to get,
\begin{equation} \label{eq:GlobalVisibility_rho2_rho0}
\frac{\rho_0^2}{\rho_2} < \left(\frac{7^3 \cdot 4^4}{3 \cdot 10^2}\right)^{1/3} \approx 6.6395.
\end{equation}
Using the above inequality, in Fig. \ref{fig:Global_Visibility_Figure}, we partition the $\rho_0-\rho_2$ parameter space into regions that develop black holes and globally visible naked singularities at the end of gravitational collapse in blue and white respectively, to obtain insight into the extent of the stability of the formation process of a Schwarzschild black hole from the gravitational collapse of a marginally bound, spherically symmetric ball of dust.

Equivalently, if we impose the condition that the initial density of the dust cloud vanishes at its boundary i.e., $\rho(0, r_b) = 0$, then we obtain $r_b = \sqrt{\rho_0/\rho_2}$ for such a cloud. We can now equivalently parametrize the space of allowed initial data using the ADM mass $M$ of the cloud and its initial radial size $r_b$. In terms of what we call the initial compactness parameter $\chi$ for this cloud,
\begin{equation}
\chi \equiv \frac{M}{r_b} = \frac{1}{15}\frac{\rho_0^2}{\rho_2},
\end{equation}
the above global visibility condition (\ref{eq:GlobalVisibility_rho2_rho0}) can be rewritten as,
\begin{equation}
\chi \lesssim 0.4426.
\end{equation}
Thus, for a marginally bound collapsing dust cloud with the density profile $\rho(0,r) = \rho_0 - \rho_2 r^2$, the cloud must start off sufficiently extended i.e., it must have a small enough mass to radius ratio (or low compactness) in order to form a global naked singularity. If the cloud is more compact than the above limit, the collapse still leads to a naked singularity, but it is of the local variety. However, when the cloud is perfectly homogeneous at the initial epoch ($\rho_2 = 0$), then the eventual singularity is not visible even locally. If we introduce the Schwarzschild (or gravitational) radius of a cloud of total mass $M$ as $r_{\text{Schw}} = 2M$, then the above equation can also be written as,
\begin{equation}
1.1297~r_{\text{Schw}} \lesssim r_b.
\end{equation}
That is, for these models, if initial radius of the dust cloud is larger than about $1.1297$ times its Schwarzschild radius, it ends up forming a globally visible naked singularity. Fig.  \ref{fig:Global_Visibility_Figure} indicates the sensitivity of the nature of the eventual singularity, that forms in marginally bound dust collapse, on initial data, i.e. it is demonstrative of the size of the basin of attraction of the Schwarzschild family of spacetime metrics.

\section{Conclusions}
We discussed the stability of equilibria of dynamical systems, in both classical mechanics and general relativity, in the framework of symplectic geometry, and attempt to set up a neat analogy to enable a simple pedagogical discussion of the notions of the stability of a spacetime. We review the Hamiltonian formulation of GR to remember how the governing equations of motion of a Hamiltonian dynamical system are simply the flow equations of the associated symplectic Hamiltonian vector field, defined on phase space. However, since here we have not accounted for gauge degeneracies in the ADM phase space, and the construction of the reduced phase space by forming the quotient space of the constrained phase space with the gauge orbits is not addressed, our results presented here are partial. The eventual goal will be to draw a formal analogy between the notions of stability in classical mechanics to those in general relativity, using its reduced phase space; the non-linear stability analysis of its critical points would simply have to do with the divergence of its flow on reduced phase space. Further, the linear stability of a critical point is concerned with the divergence of the flow of the linearization of the Hamiltonian vector field, otherwise called the tangent flow, at the critical point.

As discussed here, only the set of globally hyperbolic spacetimes can be studied within the Hamiltonian formulation of GR \cite{Inglima12}, and for such spacetimes one can apply methods of symplectic geometry. Further, there is an isomorphism from the space of all globally hyperbolic solutions of the vacuum Einstein equations to the space of allowed initial data. This is clear heuristically if one thinks of the solutions or 4-dimensional metrics $g$ as being given equivalently by a one-parameter family of 3-dimensional Riemannian metrics $q(t)$ that satisfy the flow equations of the ADM+matter symplectic Hamiltonian vector field (along with the lapse function and the shift vector, of course). Then, if one quotients out the gauge orbits, roughly a collection of all allowed $q(t)$ would correspond to the space of solutions of the Einstein equations. Since within the class of globally hyperbolic spacetimes, a particular evolution from a particular initial data set depends continuously on it, and trajectories in phase space don't intersect, it is possible to characterize a solution uniquely by its initial data $q(0)$. Typically, both these spaces (of solutions $q(t)$ and of initial data $q(0)$) are equivalent characterisations of the phase space of a physical theory and this is a feature of most typical (deterministic) physical theories like classical mechanics, quantum mechanics etc. Therefore, studying the stability of a given spacetime $g$ (orbital stability) is equivalent to studying the stability (divergence) of the Hamiltonian flow near the initial data $q(0)$ that it evolves from. Further, if one shows that a particular critical point $q_\star$, corresponding to a stationary solution $q(t) = q_\star$, is an attractor with a basin of attraction in phase space, then $q_\star$ corresponds to the metric of a linearly stable stationary spacetime. Additionally, the extent of its basin of attraction determines how non-linearly stable it is.

Since this is a nascent study, we have not analysed the potential benefits, from a numerical standpoint, of conducting a stability analysis using symplectic geometry here. One can numerically find the critical points of a Hamiltonian system by flowing along and minimising the (normal) gradient of its Hamiltonian function. Once such critical points are identified, one could study the properties of the local flow equations of the symplectic gradient of the Hamiltonian function numerically to gain insight into the nature of the stability of these critical points. Already in other contexts, for example, in applications of the theory of chaotic kinematics to oceanographic and atmospheric sciences, condensed matter, particle, accelerator and plasma physics, and also in string theory, symplectic geometry has proven to be a useful tool \cite{Rosensteel_Rowe81, Weinstein81, Atiyah+10, Baez+10, Betancourt+14, Xie+14}.

It is worth emphasising here that in general relativity, given an exact solution, obtaining the initial data set that it evolves from is generally a task of great difficulty and the branch of gravitational collapse, for example, is concerned with these issues. One of the few well characterized solutions $q(t)$ is the collapse process to a Schwarzschild black hole, which allowed us to demonstrate a non-linear stability analysis in the context of GR in \S\ref{sec:OSD_Non-Linear}. As mentioned before, some (non-globally hyperbolic) solutions of the Einstein equations cannot even be found to depend continuously on initial data. Taking advantage of the fact that the evolution of a spherically symmetric, regular cloud of pressureless matter to a Schwarzschild black hole is known to be given by the Oppenheimer-Snyder-Datt collapse, it's initial data is well characterised, and the evolutions of nearby initial data are also well understood (determined by the Lema{\^i}tre-Tolman-Bondi collapse models), we discussed the visibility of the eventual spacetime singularity that forms in these collapse evolutions. In specific, the OSD collapse to a Schwarzschild black hole evolves from homogeneous initial data ($\rho(0,r) = \rho_0$) and we considered a 2-parameter open subset of initial data $\rho(0,r) = \rho_0 - \rho_2 r^2 (\rho_2 \geq 0)$ around it. We showed that the initial compactness $\chi = M/r_b$ of collapsing cloud (where $M$ is the total ADM mass of the cloud and $r_b$ is its initial radius) governed the nature of the singularity in these models, i.e. when $\chi \lesssim .44$, the cloud formed a globally visible naked singularity and a black hole otherwise.

\textit{Acknowledgements:} Prashant Kocherlakota thanks Rukmini Dey (ICTS-TIFR, India), Franz Pedit (University of Massachusetts Amherst, USA), Rohan Poojary (CMI, India), Madhusudhan Raman (TIFR, India), Ronak M. Soni (Stanford University, USA) and Amitabh Virmani (CMI, India) for useful comments and discussions.

\end{document}